\documentclass[12pt]{article}
\usepackage{sw20lart}

\input tcilatex
\QQQ{Language}{
British English
}

\begin{document}

\title{Self-gravitating spheres of anisotropic fluid in geodesic flow}
\author{B.V.Ivanov \\
Institute for Nuclear Research and Nuclear Energy, \\
Bulgarian Academy of Science, \\
Tzarigradsko Shausse 72, Sofia 1784, Bulgaria}
\maketitle

\begin{abstract}
The fluid models mentioned in the title are classified. All characteristics
of the fluid are expressed through a master potential, satisfying an
ordinary second order differential equation. Different constraints are
imposed on this core of relations, finding new solutions and deriving the
classical results for perfect fluids and dust as particular cases. Many
uncharged and charged anisotropic solutions, all conformally flat and some
uniform density solutions are found. A number of solutions with linear
equation among the two pressures are derived, including the case of
vanishing tangential pressure.
\end{abstract}

\section{Introduction}

The description of gravitational collapse and evolution of compact objects
under various conditions remain among the important problems of general
relativity. They are described by spherically symmetric relativistic fluid
models where the metric depends on time and radius. In general, they possess
shear, expansion and acceleration, which makes them rather hard to study. In
some physical situations, however, perfect fluid (PF) models without
acceleration or dust are a good approximation. The general dust solution,
known also as the LTB solution \cite{one}, \cite{two}, \cite{three}, \cite
{four} is one of the most exploited examples. It has three branches, which
are characteristic for all other solutions. Much later the PF solution for
the parabolic branch was found \cite{five}, \cite{six}. It was shown \cite
{seven} that the other two branches lead to the Emden-Fowler equation \cite
{eight} and some exact solutions were proposed \cite{nine}. Solutions which
obey an equation of state have constant pressure and density, which is
equivalent to the presence of cosmological constant \cite{ten}. The case of
charged dust has also attracted attention \cite{eleven}, \cite{twelve}, \cite
{thirteen}. Such solutions cannot be shear-free \cite{eleven}. Finally, the
general solution was found in mass-area coordinates \cite{fourteen}.
Recently, a charged geodesic PF solution with constant pressure was studied 
\cite{fifteen}. The LTB model has been generalized to include dissipative
fluxes \cite{sixteen}.

On the other hand, different mechanisms have been identified through the
years which create pressure anisotropy in stellar models and make the fluid
imperfect \cite{seventeen}. Such are the exotic phase transitions during
gravitational collapse, the existence of a solid core or the presence of a
type P superfluid. Viscosity may also be a source of local anisotropy as
well as the slow rotation of a fluid. It has been shown that the sum of two
PFs, two null fluids or a perfect and a null fluid may be represented by
effective anisotropic fluid models \cite{eighteen}. Recently it was pointed
out that the same is true for PF with charge, bulk and shear viscosity \cite
{nineteen}. These arguments have stimulated the study of anisotropic fluids.
Geodesic anisotropic spheres were discussed in non-comoving coordinates and
a particular solution was given \cite{twenty}. Lately the notion of
Euclidean stars was introduced and their properties were investigated with
or without radiation \cite{twone}, \cite{twtwo}. In the first case the fluid
motion is geodesic. The evolution of the shear in geodesic fluids was
studied too \cite{twthree}. Unfortunately, the extensive classification of
Einstein solutions \cite{four} includes only uncharged PF. Static charged PF
solutions have been described elsewhere \cite{twfour}

The aim of the present paper is to generalize the above results and give a
classification of the geodesic anisotropic fluid solutions, similar to that
of shear-free anisotropic fluid solutions \cite{twfive}. For this purpose we
use the reformulation of the Einstein equations in terms of the mass
function \cite{one}, \cite{twsix}. In this approach, besides the new
solutions, the classical results reviewed above follow as particular cases.

In Sec.2 the field equations based on the mass function are given for
non-radiating spheres in geodesic flow. The fundamental potential and the
main equation that it satisfies are introduced. Expressions for the other
characteristics of the metric are given. Sec.3 discusses the general and
some concrete solutions of the geodesic anisotropic model. In Sec.4 the
particular cases of PF and dust are investigated, making contact with the
existing literature. In Sec.5 the general charged solution and its subcases
of charged PF and charged dust are discussed. Sec.6 studies the case when
the radial and the tangential pressures are proportional. In Sec.7 the
general conformally flat solution for geodesic anisotropic fluids is found.
The uniform density solution is studied in Sec.8. In Sec.9 a comparison is
made between shear-free and geodesic fluids. The final section contains some
discussion.

\section{Field equations for geodesic fluids}

Spherically symmetric relativistic fluid models are described by the metric 
\begin{equation}
ds^2=-e^{2\nu }dt^2+e^{2\lambda }dr^2+R^2\left( d\theta ^2+\sin ^2\theta
d\varphi ^2\right) ,  \label{one}
\end{equation}
where $\nu ,\lambda $ and $R$ are functions of $t$ and $r$ only. The
spherical coordinates are numbered as $x^0=t,x^1=r,x^2=\theta ,x^3=\varphi $%
. The Einstein equations read 
\begin{equation}
R_{ij}-\frac 12g_{ij}R_{\;k}^k=8\pi T_{ij}.  \label{two}
\end{equation}
Here $R_{ij}$ is the Ricci tensor, $T_{ij}$ is the energy-momentum tensor
and we have set $G=c=1$. For an anisotropic fluid model without radiation
one has \cite{twseven} 
\begin{equation}
T_{ik}=\left( \rho +p_t\right) u_iu_k-p_tg_{ik}+\left( p_r-p_t\right) \chi
_i\chi _k.  \label{three}
\end{equation}
Here $\rho $ is the energy density, $p_r$ is the radial pressure, $p_t$ is
the tangential pressure, $u^i$ is the four-velocity of the fluid, $\chi ^i$
is a unit spacelike vector along the radial direction. These vectors satisfy
the relations 
\begin{equation}
u^iu_i=-1,\qquad \chi ^i\chi _i=1,\qquad u^i\chi _i=0.\qquad  \label{four}
\end{equation}
The coordinates are assumed to be comoving, hence 
\begin{equation}
u^i=e^{-\nu }\delta _0^i,\qquad \chi ^i=e^{-\lambda }\delta _1^i.
\label{five}
\end{equation}

The usual field equations (2) are rather cumbersome except for the (01)
component 
\begin{equation}
\dot R^{\prime }-\dot R\nu ^{\prime }-R^{\prime }\dot \lambda =0.
\label{six}
\end{equation}
The dot above means time derivative, while the prime denotes a radial one.
It is easier to work in the mass function formalism \cite{one}, \cite{twsix}%
, namely 
\begin{equation}
m^{\prime }=4\pi \rho R^2R^{\prime },  \label{seven}
\end{equation}
\begin{equation}
\dot m=-4\pi p_rR^2\dot R,  \label{eight}
\end{equation}
\begin{equation}
\dot \lambda \left( \rho +p_r\right) =-\dot \rho -\frac{2\dot R}R\left( \rho
+p_t\right) ,  \label{nine}
\end{equation}
\begin{equation}
\nu ^{\prime }\left( \rho +p_r\right) =-p_r^{\prime }+\frac{2R^{\prime }}%
R\left( p_t-p_r\right) .  \label{ten}
\end{equation}
The mass function is given by 
\begin{equation}
m=\frac 12R\left( e^{-2\nu }\dot R^2+k\right) ,  \label{eleven}
\end{equation}
\begin{equation}
k=1-e^{-2\lambda }R^{\prime 2},  \label{twelve}
\end{equation}
where $k$ is the binding energy \cite{tweight}. Bound configurations have $%
0<k<1$, unbound ones have $k<0$ and the marginally bound case is $k=0$.
These three cases are called also elliptic, hyperbolic and parabolic
respectively. The second Weyl invariant reads \cite{twfive}

\begin{equation}
\Psi _2=\left( \frac m{R^3}\right) ^{\prime }\frac R{3R^{\prime }}+\frac{%
4\pi }3\bigtriangleup p,\qquad \bigtriangleup p=p_t-p_r.  \label{thirteen}
\end{equation}

The expansion of the fluid is 
\begin{equation}
\Theta =e^{-\nu }\left( \dot \lambda +\frac{2\dot R}R\right) ,
\label{fourteen}
\end{equation}
while the components of the shear tensor are proportional to 
\begin{equation}
\sigma =\frac 13e^{-\nu }\left( \frac{\dot R}R-\dot \lambda \right) .
\label{fifteen}
\end{equation}
The four acceleration is $\dot u_i=\left( 0,\nu ^{\prime },0,0\right) $.

Geodesic fluids have vanishing acceleration, hence $\nu ^{\prime }=0$ and $%
\nu $ is a function of time only. It can be set to zero by a coordinate
transformation. If $k=0$ Eqs (6) and (12) lead to geodesic flow. The
contrary is not true. When $\nu =0$ and $R^{\prime }\neq 0$ the same
equations yield $k=k\left( r\right) $ and 
\begin{equation}
e^{2\lambda }=\frac{R^{\prime 2}}{1-k}.  \label{sixteen}
\end{equation}
It is not possible to make $k$ vanish by a transformation of $r$. The so
called Euclidean star models \cite{twone} have $k=0$. Eq (8) becomes with
the help of Eq (11) 
\begin{equation}
\left( R\dot R^2\right) ^{.}=-8\pi p_rR^2\dot R-k\dot R.  \label{seventeen}
\end{equation}
This is the main equation for geodesic fluids. It is second order and
contains only time derivatives. The dependence on $r$ is viewed upon as
parametric. One gets rid of the first derivatives by the substitution 
\begin{equation}
R=Z^{2/3},  \label{eighteen}
\end{equation}
which results in 
\begin{equation}
\ddot Z=-6\pi p_rZ-\frac 34kZ^{-1/3}.  \label{nineteen}
\end{equation}
We can find $Z$ when $p_r$ is given or vice versa. The second case is much
simpler but it is not guaranteed that the resulting $p_r$ will be physically
realistic. Thus the general solution depends on one arbitrary function of $t$
and $r$.

Eq (10) becomes 
\begin{equation}
p_r^{\prime }=\frac{2R^{\prime }}R\left( p_t-p_r\right)  \label{twenty}
\end{equation}
and gives an expression for the tangential pressure 
\begin{equation}
p_t=\frac{\left( R^2p_r\right) ^{\prime }}{\left( R^2\right) ^{\prime }}.
\label{twone}
\end{equation}
When $p_r=0$, $p_t$ vanishes too and we get a dust solution, but the
opposite is not true. The mass is found from Eq (11), which becomes 
\begin{equation}
m=\frac 12R\left( \dot R^2+k\right) =\frac 29\dot Z^2+\frac k2Z^{2/3}.
\label{twtwo}
\end{equation}
Then the energy density follows from Eq (7) 
\begin{equation}
24\pi \rho =\frac{4\left( \dot Z^2\right) ^{\prime }+9\left( kZ^{2/3}\right)
^{\prime }}{\left( Z^2\right) ^{\prime }}.  \label{twthree}
\end{equation}
The expansion and the shear scalar are given by 
\begin{equation}
\Theta =\frac{\left( Z^2\right) ^{\prime .}}{\left( Z^2\right) ^{\prime }},
\label{twfour}
\end{equation}
\begin{equation}
\sigma =-\frac{\left( \ln Z\right) ^{\prime .}}{3\left( \ln Z\right)
^{\prime }}  \label{twfive}
\end{equation}
A geodesic solution is also shear-free when $Z$ and consequently $R$ is
separable. It has no expansion when $R$ is of the form 
\begin{equation}
R\left( t,r\right) =\left[ R_1\left( t\right) +R_2\left( r\right) \right]
^{1/3}.  \label{twsix}
\end{equation}

\section{General and particular solutions}

The main equation (19) may be simplified further by the transformation 
\begin{equation}
Z=\mu W,\qquad \frac{dt}{d\tau }=\mu ^2,  \label{twseven}
\end{equation}
where $\mu \left( t,r\right) $ determines the new time variable $\tau $. It
becomes 
\begin{equation}
W_{\tau \tau }=-\left( \ddot \mu +6\pi p_r\mu \right) \mu ^3W-\frac 34k\mu
^{8/3}W^{-1/3}.  \label{tweight}
\end{equation}
We choose $\mu $ so that the bracket vanishes and the above equation breaks
into two 
\begin{equation}
6\pi p_r=-\frac{\ddot \mu }\mu =\frac{2\mu _\tau ^2-\mu \mu _{\tau \tau }}{%
\mu ^5},  \label{twnine}
\end{equation}
\begin{equation}
W_{\tau \tau }=-\frac 34k\mu ^{8/3}W^{-1/3}.  \label{thirty}
\end{equation}

The general solution follows when one chooses arbitrary $k\left( r\right) $, 
$\mu \left( t,r\right) $ or $p_r\left( t,r\right) $, and finds $W\left( \tau
,r\right) $ from the above equation. This determines consecutively $Z\left(
t,r\right) $, $R\left( t,r\right) $ and all other characteristics of the
fluid. Eq (30) for $k\neq 0$ resembles the Emden-Fowler (EF) equation \cite
{eight} 
\begin{equation}
W_{\tau \tau }=A\tau ^nW^l  \label{thone}
\end{equation}
with $A=-\frac 34k$ and $l=-1/3$. In order to obtain analytical solutions we
set $\mu =\tau ^{3n/8}$. Eq (27) yields for $n\neq -4/3$ and $n=-4/3$
respectively 
\begin{equation}
\tau ^{\frac{3n+4}4}=\frac{3n+4}4\left( t-a\right) ,\qquad \tau =e^{t-a},
\label{thtwo}
\end{equation}
where $a\left( r\right) $ is an arbitrary function of integration. Eq (29)
gives for $n\neq -4/3$ and $n=-4/3$ respectively 
\begin{equation}
6\pi p_r=\frac{3n\left( 3n+8\right) }{4\left( 3n+4\right) ^2\left(
t-a\right) ^2},\qquad 6\pi p_r=-\frac 14.  \label{ththree}
\end{equation}

There are general solutions of Eq (30) when $n=0$ ($\mu =1$) or $n=-8/3$ ($%
\mu =1/\tau $) \cite{eight}, but according to Eq (29) they are particular
dust solutions, which are discussed in the next section. The case $n=-4/3$ ($%
\mu =\tau ^{-1/2}$) is also soluble, but leads to constant pressures,
representing effectively a cosmological constant.

There is, however, a particular two-parameter solution of Eq (31) for any $%
n,l\neq 1$%
\begin{equation}
W=\alpha \tau ^{\frac{n+2}{1-l}},\qquad \alpha =\left[ \frac{\left(
n+2\right) \left( n+l+1\right) }{A\left( l-1\right) ^2}\right] ^{\frac
1{l-1}}.  \label{thfour}
\end{equation}
In our case $l$ is fixed, while $n$ may be a function of $r$. We obtain 
\begin{equation}
R=\alpha _1\left( t-a\right) ,\qquad \alpha _1^2=-\frac{\left( 3n+4\right)
^2k}{4\left( n+2\right) \left( 3n+2\right) }.  \label{thfive}
\end{equation}
The solution depends on three arbitrary functions of the radial coordinate $%
k,a,n$. When $a=0$, $R$ is separable, hence, the solution is shear-free in
addition to being geodesic.

No other explicit solutions exist when $\mu $ is a power of $\tau $.
However, there is one with $\mu =\left( a\tau ^2+b\tau +c\right) ^{-1/2}$
where $a,b,c$ are functions of $r$ . It leads to time-independent radial
pressure $p_r\left( r\right) $. The pressure profile is given by the initial
conditions and may be arbitrary. This case is better approached from Eq
(19), which is integrated simply by multiplication with $\dot Z$. We get the
integral 
\begin{equation}
\int \frac{dZ}{\sqrt{-6\pi p_rZ^2-\frac 94kZ^{2/3}+h}}=t+s,  \label{thsix}
\end{equation}
where $h\left( r\right) ,s\left( r\right) $ are arbitrary integration
functions. The first is related to the mass functions through Eq (22) 
\begin{equation}
m=-\frac 43\pi p_rR^3+\frac 29h.  \label{thseven}
\end{equation}
The integral in Eq (36) has an analytical expression when any of the
functions $p_r,k,h$ vanishes. Thus, when $h=0$ we resort back to $R$ to find 
\begin{equation}
R=\left( \frac{-3k}{8\pi p_r}\right) ^{1/2}\sin \frac 23\sqrt{6\pi p_r}%
\left( t+s\right) ,  \label{theight}
\end{equation}
provided $p_r>0$ and $k<0$. When $k=0$ we obtain in a similar way 
\begin{equation}
R^{3/2}=\sqrt{\frac h{6\pi p_r}}\sin \sqrt{6\pi p_r}\left( t+s\right) .
\label{thnine}
\end{equation}
This is a concrete example of an Euclidean star. Finally, when $p_r=0$ a
dust solution follows, to be discussed in the next section.

Up to now we have assumed that $k\neq 0$ in Eq (30). The case $k=0$ gives
immediately 
\begin{equation}
W=a\tau +b,\qquad \tau =\int \mu ^{-2}dt,  \label{forty}
\end{equation}
\begin{equation}
R=\mu ^{2/3}\left( a\int \mu ^{-2}dt+b\right) ^{2/3},  \label{foone}
\end{equation}
where $a\left( r\right) ,b\left( r\right) $ and $\mu \left( t,r\right) $ are
arbitrary. The radial pressure is found from Eq (29). When $b=0$, $R$
becomes separable and the solution is shear-free too.

\section{Perfect fluid and dust}

Perfect fluid may be considered as anisotropic fluid with the equation of
state (isotropy condition) $p_r=p_t\equiv p$. In this case Eq (20) shows
that $p=p\left( t\right) $. It is enough to take a $\mu \left( t\right) $ in
Eq (29). In fact, the transformation given by Eq (27) was applied first to
PF \cite{seven}, \cite{nine}. The solution (36) was also proposed \cite{nine}%
, without noticing that it holds for $p\left( r\right) $ and contradicts the
PF condition. The concrete solution for $k\neq 0$ described by Eqs (33, 35)
holds for PF too, provided $a,n$ are constant. The general solution in the
case $k=0$ is given again by Eq (41), but with $\mu $ uniform in space \cite
{five}, \cite{six}. Szafron \cite{twnine} has found the particular solution $%
\mu =t^q$ with $q$ any real number.

When the pressures vanish, one has collapsing dust. Eq (10) shows that its
motion is geodesic (without acceleration). According to Eq (8) the mass
function is time-independent. Its profile characterizes the solution. The
main equation (19) may be integrated once and becomes Eq (22). A second
integration leads to Eq (36) with $p_r=0$ and $h=9m/2$ , in accord with Eq
(37). This is the well known LTB solution \cite{one}, \cite{two}, \cite
{three}. Eq (22) coincides with Eq (73) from Ref. \cite{sixteen}, where one
can find the analytic expressions for $R$ in the parabolic, hyperbolic and
elliptic case. For example, when $k=0$ one easily finds 
\begin{equation}
R=\left( 3\sqrt{\frac m2}t+b\right) ^{2/3},  \label{fotwo}
\end{equation}
$m\left( r\right) ,b\left( r\right) $ being arbitrary functions. The
solution is also shear-free when $b=0$.

\section{Charged fluid}

Spherical symmetry allows the appearance of only an electric field $E$ in
the radial direction. The energy-momentum tensor of this field can be
described \cite{nineteen} as addition of the following effective energy
density and pressures to the fluid 
\begin{equation}
\rho ^E=p_t^E=-p_r^E=\frac{E^2}{8\pi }.  \label{fothree}
\end{equation}
The Maxwell equations give 
\begin{equation}
\frac{E^2}{8\pi }=\frac{q\left( r\right) }{R^4},\qquad 4\pi C=Ee^{-\lambda },
\label{fofour}
\end{equation}
where $C$ is the charge function of the fluid and $q\left( r\right) $ is an
arbitrary function. The main equation in its two forms (19) and (30) becomes 
\begin{equation}
\ddot Z=-6\pi p_rZ-\frac 34kZ^{-1/3}+6\pi qZ^{-5/3},  \label{fofive}
\end{equation}
\begin{equation}
W_{\tau \tau }=-\frac 34k\mu ^{8/3}W^{-1/3}+6\pi q\mu ^{4/3}W^{-5/3},
\label{fosix}
\end{equation}
while Eqs (27) and (29) do not change. Eq (21) transforms into 
\begin{equation}
p_t=\frac{\left( R^2p_r\right) ^{\prime }}{\left( R^2\right) ^{\prime }}-%
\frac{q^{\prime }}{2R^3R^{\prime }}  \label{foseven}
\end{equation}
Now when $p_r$ vanishes $p_t$ is not obliged to vanish, giving a class of
non-dust solutions.

For $k\neq 0$ Eq (46) becomes the modified EF equation 
\begin{equation}
W_{\tau \tau }=A_1\tau ^{n_1}W^{m_1}+A_2\tau ^{n_2}W^{m_2}  \label{foeight}
\end{equation}
when $\mu =\tau ^{3n_1/8}$. Then 
\begin{equation}
m_1=-\frac 13,\quad m_2=-\frac 53,\quad n_1=2n_2,\quad A_1=-\frac 34k,\quad
A_2=6\pi q.  \label{fonine}
\end{equation}

There is a table of 108 explicit solutions of Eq (48) \cite{eight}. Those
which satisfy the conditions above have $n_1=0;-\frac 83;-\frac 43$. The
first two lead to vanishing $p_r$, the third has it constant. They are
non-trivial and non-dust. They can be approached also as particular cases of
the class of solutions with $p_r\left( r\right) $. Like for uncharged fluid
one integrates directly Eq (45) to find 
\begin{equation}
\frac 32\int \frac{RdR}{\sqrt{-6\pi p_rR^4-\frac 94kR^2+hR-16\pi q}}%
=t+s\left( r\right) .  \label{fifty}
\end{equation}
Eq (50) coincides with Eq (36) if $q=0$. When $h$ or $p_r$ vanishes, the
integral is expressed in terms of elementary functions in five different
ways depending on the relations between $k,q$ and $p_r$ or $h$.

For $k=0,$ $A_1=0$ and the first term in Eqs (46, 48) disappears. We get
once more the usual EF equation like Eqs (30, 31) in the uncharged case, but
the power of $W$ is different. This change brings a host of explicit
solutions with $n_2=0;-\frac 43;-\frac 23;-\frac{10}3;-\frac 73;-\frac
56;-\frac 12;1;2$. The first two have vanishing radial pressure. The
particular solution (34) also holds provided $l=-5/3$ $.$

Charged PF has $p_r=p_t\equiv p$. Eq (47) becomes 
\begin{equation}
p^{\prime }=\frac{q^{\prime }}{R^4}.  \label{fione}
\end{equation}
This is another relation between $p$ and $R$ in addition to Eq (45). A
possible solution is $p\left( t\right) $ (like in the uncharged case) and $%
q=const$. Then $\mu =\mu \left( t\right) $ and $a,n$ should be constant in
the above solutions. The isotropic pressure is given by Eq (33).

When $p^{\prime }\neq 0$ one should replace $R$ from Eq (51) into Eq (46)
having in mind that 
\begin{equation}
W=\mu ^{-1}R^{3/2}.  \label{fitwo}
\end{equation}
Even in the simplest case $k=0$ this leads to contradiction, because the
result is $n=-4/3$ and hence, $p=const$.

Charged dust may be considered as a subclass of the PF solutions with $p=0$.
The general solution is known, though in another coordinate system \cite
{fourteen}. However, Eq (10) does not lead now to $\nu ^{\prime }=0$ because
of the electromagnetic additions to the energy density and the pressures
given by Eq (43). We have to impose this condition in order to make the
fluid flow geodesic and the solutions found here form a subclass of the
general Ori solution. When $p=0$ Eq (51) yields $q\left( r\right) =q_0=const$%
. It is not necessary to introduce $W$ and $\mu $. Eq (45) becomes 
\begin{equation}
\ddot Z=-\frac 34kZ^{-1/3}+6\pi q_0Z^{-5/3}.  \label{fithree}
\end{equation}
This is the modified EF equation (48, 49) with $n_1=n_2=0$ and $A_2=6\pi
q_0. $ Its solution is given by example 2.4.2.56 from \cite{eight} and has
three branches.

When $k<0$%
\begin{equation}
t=C_1e^{\omega \lambda }+C_2e^{-\omega \lambda }+C_3\lambda ,\qquad R=\omega
\left( C_1e^{\omega \lambda }-C_2e^{-\omega \lambda }\right) +C_3,
\label{fifour}
\end{equation}
where $\lambda $ is a parameter, $\omega =\sqrt{|k|}$ and the integration
functions $C_i\left( r\right) $ satisfy the relation 
\begin{equation}
3\left( A_1C_3^2+A_2\right) +16A_1^2C_1C_2=0.  \label{fifive}
\end{equation}

When $k>0$%
\begin{equation}
t=C_1\sin \omega \lambda +C_2\cos \omega \lambda +C_3\lambda ,\quad R=\omega
\left( C_1\cos \omega \lambda -C_2\sin \omega \lambda \right) +C_3,
\label{fisix}
\end{equation}
\begin{equation}
3\left( A_1C_3^2+A_2\right) +4A_1^2\left( C_1^2+C_2^2\right) =0.
\label{fiseven}
\end{equation}

Finally, when $k=0$ Eq (53) may be integrated by multiplying with $\dot Z$.
The result is 
\begin{equation}
\int \frac{RdR}Y=\frac 23t+C_2,  \label{fieight}
\end{equation}
where 
\begin{equation}
Y=\left( C_1R-18\pi q_0\right) ^{1/2}.  \label{finine}
\end{equation}
Eq (58) becomes a cubic equation for $Y$%
\begin{equation}
Y^3+54\pi q_0Y=C_1^2\left( t+\frac 32C_2\right) .  \label{sixty}
\end{equation}
It may be solved for $Y$ and $R$. The solutions are not separable and
therefore are not shear-free in accord with Ref. \cite{eleven}.

\section{Solutions with $p_t=\gamma p_r$}

This equation of state among the pressures includes the cases of perfect
fluid and $p_t=0$. Eq (21), which in principle determines the tangential
pressure, now gives an expression for $p_r$ in terms of $Z$%
\begin{equation}
p_r=h\left( t\right) Z^{\frac 43\left( \gamma -1\right) },  \label{sione}
\end{equation}
where $h$ is an arbitrary positive function. The main equation (19) becomes 
\begin{equation}
\ddot Z=-\frac 34k\left( r\right) Z^{-1/3}-6\pi h\left( t\right) Z^{\frac{%
4\gamma -1}3}.  \label{sitwo}
\end{equation}
This represents the modified EF equation (48) with 
\begin{equation}
m_1=-1/3,\quad m_2=\frac{4\gamma -1}3,\quad n_1=0,\quad A_1=-\frac 34k,\quad
A_2=-6\pi ,  \label{sithree}
\end{equation}
provided $h=t^{n_2}$. The following explicit solutions exist:

When $k\neq 0$ we may take $n_2=0,m_2=-5/3$ ($h=h_0,\gamma =-1$); then Eq
(62) coincides in form with Eq (53) and the solution (54-57) holds. Other
possibilities with the same $m_2$ are $n_2=1$ and $n_2=2$ which lead to
examples 2.4.2.58 and 2.4.2.54 respectively \cite{eight}. The case $p_t=0$ ($%
\gamma =0$) reduces Eq (62) to 
\begin{equation}
\ddot Z=-\left( \frac 34k+6\pi h\right) Z^{-1/3}.  \label{sifour}
\end{equation}
It is easily integrable for constant $h=h_0$ when it coincides in form with
the LTB solution.

When $k=0$ Eq (62) becomes the usual EF equation with $m_1=\frac{4\gamma -1}3
$ provided $h=t^{n_1}$. One can choose from the 28 analytic solutions (5
one-parameter families and 23 isolated points) \cite{eight} and the
particular solution (34).

Eqs (21, 61) show that when $\gamma <1$ (including the case of vanishing
tangential pressure) the radial pressure becomes infinite at $R=0$, adding
another singularity to the curvature singularity of the LTB solution. This
is true provided there is no bounce of the collapsing fluid and the point $%
R=0$ is covered by it after some time. A similar situation is described in
Ref. \cite{tweight}.

\section{Conformally flat solutions}

These solutions have $\Psi _2=0.$ Combining Eqs (13, 20) yields for geodesic
fluids 
\begin{equation}
\Psi _2=\left( \frac m{R^3}+2\pi p_r\right) ^{\prime }\frac R{3R^{\prime }}
\label{sifive}
\end{equation}
The vanishing of the second Weyl invariant requires 
\begin{equation}
\frac m{R^3}+2\pi p_r=f\left( t\right) ,  \label{sisix}
\end{equation}
where $f$ is an arbitrary positive function since $m$ and $p_r$ are positive
for realistic star models. Eq (8) transforms into 
\begin{equation}
\frac{\dot X}{2f+X}=-\frac{\dot R}R,\qquad X\equiv f-2\pi p_r.
\label{siseven}
\end{equation}
Inserting Eq (22) into Eq (66) results in an expression for $p_r$%
\begin{equation}
2\pi p_r=f-\frac{\dot R^2+k}{2R^2}.  \label{sieight}
\end{equation}
We combine this equation with Eq (66) to find 
\begin{equation}
2R\ddot R-\dot R^2=-4fR^2+k,  \label{sinine}
\end{equation}
which is analogous to Eq (17). The first derivative is eliminated by the
substitution $R=Y^2$%
\begin{equation}
\ddot Y=-fY+\frac k4Y^{-3}.  \label{seventy}
\end{equation}
This is the main equation in the conformally flat case, replacing Eq (19).
It has no reference to the pressure, although $f$ enters it in a similar
way. Once again we use the transformation $Y=\mu W$ which splits Eq (70)
into two parts 
\begin{equation}
\ddot \mu +f\mu =0,\qquad W_{\tau \tau }=\frac k4W^{-3},  \label{seone}
\end{equation}
where $\mu =\mu \left( t\right) $. Once again we obtain a EF equation. This
time it is easily integrated by multiplying with $W_\tau $. We get 
\begin{equation}
R=b\mu ^2\left( \frac 14V^2+\frac k{b^2}\right) ,\qquad V\equiv \int \mu
^{-2}dt+c.  \label{setwo}
\end{equation}
Here $b\left( r\right) ,c\left( r\right) $ are arbitrary integration
functions and $\mu ^2$ is an arbitrary positive function of time. This
formula is valid for any $k$. In the marginal case $f=0$ (which requires
some radial tension) it simplifies to 
\begin{equation}
R=\frac b4\left( t+c\right) ^2+\frac kb.  \label{sethree}
\end{equation}
The general conformally flat geodesic solution is also shear-free provided $%
c=0=k$.

Let us find now the conformally flat PF solution. Eq (20) shows that $%
p=p\left( t\right) $. Eq (68) becomes 
\begin{equation}
\frac{\dot R^2+k}{2R^2}=X\left( t\right) .  \label{sefour}
\end{equation}
Eq (22) then gives for the mass function 
\begin{equation}
m=\frac 12X\left( t\right) R^3  \label{sefive}
\end{equation}
and Eq (7) shows that the energy density is uniform, $\rho =\rho \left(
t\right) $. Putting the general solution (72) into Eq (74) yields 
\begin{equation}
\frac{1+2\mu \dot \mu V}{\frac 14V^2+\frac k{b^2}}=X_1\left( t\right) .
\label{sesix}
\end{equation}
The l.h.s. should depend only on time, which is ensured by setting to
constants $c=c_0$ and $k/b^2=c_1$. $R$ becomes 
\begin{equation}
R=b\mu ^2\left[ \frac 14\left( \int \mu ^{-2}dt+c_0\right) ^2+c_1\right] .
\label{seseven}
\end{equation}
The solution is shear-free because $R$ is separable.

The conformally flat dust solution is a subcase of the PF solution with
vanishing pressure. Eqs (67, 68) become 
\begin{equation}
\frac{\dot f}{3f}=-\frac{\dot R}R,\qquad \dot R^2+k=2fR^2.  \label{seeight}
\end{equation}
They lead to 
\begin{equation}
R=g\left( r\right) f\left( t\right) ^{-1/3},  \label{senine}
\end{equation}
\begin{equation}
\dot f=3f^{4/3}\sqrt{2f^{1/3}-\frac k{g^2}},  \label{eighty}
\end{equation}
where $g\left( r\right) $ is arbitrary, but positive. Obviously $%
k/g^2=c_2=const$. The equation is integrated by the substitution $y=f^{1/3}$
and the final result is 
\begin{equation}
R=\left( \frac 3{\sqrt{2}}\right) ^{2/3}g\left( r\right) \left( t+c_3\right)
^{2/3},  \label{eione}
\end{equation}
where $c_3$ is some constant. This formula resembles the $k=0$ case of the
LTB solution given by Eq (42). Like the PF solution (77), the dust solution
is shear-free.

\section{Uniform density}

The energy density in this case is a function of time, $\rho \left( t\right) 
$. Eq (7) gives upon integration 
\begin{equation}
m=\frac{4\pi }3\rho R^3,  \label{eitwo}
\end{equation}
where an integration function has been set to zero \cite{thirty}. Plugging
this formula in Eq (22) yields 
\begin{equation}
\frac{8\pi }3\rho =\frac{\dot R^2+k}{R^2}.  \label{eithree}
\end{equation}
With the help of Eq (8) it is transformed into 
\begin{equation}
3k=R^2\left[ 8\pi \rho -\frac{\dot \rho ^2}{3\left( \rho +p_r\right) ^2}%
\right] .  \label{eifour}
\end{equation}
If we put Eq (82) into Eq (8) an expression for the radial pressure is
obtained 
\begin{equation}
-4\pi p_r=\frac{\ddot R}R+\frac{4\pi }3\rho .  \label{eifive}
\end{equation}

Let us discuss first the case $k=0$. Eq (84) defines $p_r$ in terms of the
arbitrary $\rho \left( t\right) $%
\begin{equation}
p_r=\pm \frac{\dot \rho }{\sqrt{24\pi \rho }}-\rho .  \label{eisix}
\end{equation}
It shows that $p_r=p_r\left( t\right) $ and the fluid is perfect. Eq (83)
defines $R$%
\begin{equation}
R=a\left( r\right) \exp \left( \pm \sqrt{\frac{8\pi }3}\int \sqrt{\rho }%
dt\right)  \label{eiseven}
\end{equation}
with integration function $a\left( r\right) $. The solution is shear-free.

In the dust subcase Eq (86) becomes an equation for $\rho .$ Its solution is 
\begin{equation}
\rho =\frac 1{6\pi \left( t+t_0\right) ^2},  \label{eieight}
\end{equation}
where $t_0$ is an integration constant. Then $R$ simplifies 
\begin{equation}
R=a\left( r\right) \left( t+t_0\right) ^{2/3}.  \label{einine}
\end{equation}
It coincides in form with the conformally flat dust solution given by Eq
(81). However the function $g\left( r\right) $ there depends on $k$ and it
holds for any $k$.

Let us consider now the case $k\neq 0$. We set 
\begin{equation}
R\left( t,r\right) =\sqrt{|k\left( r\right) |}P\left( t,r\right) .
\label{ninety}
\end{equation}
Eq (83) becomes 
\begin{equation}
\dot P^2=\frac{8\pi }3\rho P^2-\varepsilon ,\qquad \varepsilon \equiv
sign\,k.  \label{nione}
\end{equation}
This equation is not integrable in general and separable solutions are not
possible. For PF, however, Eqs (84, 90) show that $P=P\left( t\right) $ and
we may take it as an arbitrary function which determines $\rho $ and $R$.
The pressure is found from Eq (85). In the dust subcase $m=m\left( r\right) $
and Eq (82) becomes 
\begin{equation}
\rho P^3=\frac{3m}{4\pi |k|^{3/2}}.  \label{nitwo}
\end{equation}
The l.h.s. depends on $t$, while the r.h.s. depends on $r$, hence, both of
them are constant, $3c_0/8\pi $. Then Eq (91) reads 
\begin{equation}
\dot P^2=\frac{c_0-\varepsilon P}P  \label{nithree}
\end{equation}
and may be integrated, giving a rather complicated inexplicit expression for 
$P$%
\begin{equation}
-\sqrt{P\left( c_0-\varepsilon P\right) }+\frac 12\ln \left| \frac
P{P-\varepsilon c_0}\right| =\varepsilon \left( t+t_0\right) .
\label{nifour}
\end{equation}
It is not possible to recover the $k=0$ case by setting $\varepsilon =0$.

\section{Comparison with shear-free anisotropic fluid}

The structure of the present classification of geodesic anisotropic fluid
spheres is analogous to that of shear-free anisotropic fluid spheres \cite
{twfive}, denoted further as I. In both cases the (01) component of the
Einstein equations may be integrated. As a result all metric components
depend on $R$ plus arbitrary functions of one variable like $\Theta \left(
t\right) $ in I and $k\left( r\right) $ here. The latter function divides
any solution into two or three branches, namely $k=0$ and $k\neq 0$ ($k>0$
and $k<0$ in some cases).

In both cases the fundamental ingredient is a second order differential
equation for $R$ in only one of the variables $t$ or $r$. It is based on the
expression for the mass function, which luckily contains derivatives of only
one kind. For shear-free fluids this is Eq (I 34) with only first order
derivatives in $r$. It leads to the main Eq (I 36), which is of second
order. For geodesic fluids the mass function is given by Eq (22), containing
only a time derivative. It leads to the main Eq (19), which is of second
order in time derivatives. All other characteristics of the fluid are
expressed through the solutions of the main equation

Shear, expansion and acceleration are usually expressed through the metric.
An essential role in the formalism plays a second formula for the vanishing
characteristic, which holds for any fluid sphere. It elevates its dependence
from metric components to sources (pressures and energy density). In I it is
given by Eq (I 29) for the shear, whose derivation required the mass
decomposition formula (I 15). Here this is Eq (10) for the
four-acceleration, which enters the usual set of Einstein equations, based
on the mass function. It is not necessary to use the mass decomposition
formula except for the definition of $\Psi _2$ given by Eq (13).

The main equation contains no first derivatives and just one non-derivative
term, when one passes from $R$ to $L=r/R$ in I or to $Z=R^{3/2}$ here. For
geodesic fluids a second step is necessary, namely the transformation in Eq
(27). The general solution of the main solution is based on one arbitrary
function. In I this may be $R$, $L$, $Z$, $\bigtriangleup p$ or $\Psi _2.$
Here these are $R$, $Z$ or $p_r$. When $R$ is chosen, solutions are found
quite easily, however, they may be unphysical and the passage to PF is
rather involved. Therefore we take $Z$ or $\bigtriangleup p$ as arbitrary
functions in I or $p_r$ here and solve the main equation. Analytic solutions
of it are found by reduction to an EF equation. This happens when an
arbitrary function is taken as a power of $r^2$ in I or $\tau $ in the
present paper.

The passage to PF is controlled by $\bigtriangleup p$ in I (it enters
directly Eq (I 36)) and by $p_r^{\prime }=0$ here, which makes the radial
pressure dependent on time only. For geodesic fluids the additional case of
dust exists. It is treated as a subcase of PF, when the pressure vanishes.

Charged fluid has additions to the pressures and density which bring a
second term to the main equation. It becomes the modified EF equation and
possesses a number of one-parameter or isolated solutions.

There is one easily integrable charged case when some arbitrary functions
are set to constants for shear-free fluids or $p_r=p_r\left( r\right) $ for
geodesic fluids. The corresponding integral (I 52) leads to the Weierstrass
elliptic function in I and to Eq (50) here, which has analytic expression in
some cases. In I an equation of state for PF emerges from this case. Here
the necessary condition for this is constant pressure, which represents a
cosmological constant in accord with Ref.\cite{ten} and is rather trivial.
The case of charged dust is completely integrable.

The general conformally flat solution is found in both papers and is given
by Eq (I 64) and Eq (72) respectively. The solution for PF follows when an
arbitrary function depending on $r$ is put to zero in I, or two such
functions are made constant for geodesic fluids. Conformally flat dust is
obtained by further simplification. Conformally flat PF solutions intertwine
with uniform density solutions.

The general uniform density solution is found in I, while here this is done
in the $k=0$ case for anisotropic fluid and in all cases for PF and dust.

An important case is to impose a linear equation of state $p_t=\gamma p_r$
between the two pressures, which is a generalization of the isotropy
condition $\gamma =1$, leading to PF. In both cases there are analytic
solutions for $R$ at some discrete values of $\gamma $, including the case
of vanishing tangential pressure $\gamma =0$.

A geodesic solution may be shear-free too. This happens when $R$ is
separable. Many of the above solutions become separable when the integration
functions are chosen appropriately. However, we were unable to find
solutions that are both geodesic and expansion-free, i.e. satisfy Eq (26),
except for the trivial case $R\left( r\right) $.

As a final remark, the EF equation or its modified version appear frequently
in both the shear-free and the acceleration-free case. Integrals which lead
to the Weierstrass function are as frequent in I but here they do not appear
at all.

\section{Discussion}

Recently we have stressed the importance of anisotropic fluid models for
astrophysics \cite{nineteen}. The fluid flow has three important
characteristics - shear, acceleration and expansion. It is hard to obtain
general solutions when all three are non-trivial. We have classified in
another paper the shear-free anisotropic spheres \cite{twfive}. In the
present paper the same is done for geodesic flows, where the acceleration
vanishes.

Taking the fluid anisotropic means accepting the maximum freedom allowed by
spherical symmetry. The general solution depends on an arbitrary function
and a core of simple relations is obtained. Only after that we start to
impose different constraints aiming the system at particular cases.

Thus, PF is this core with imposed isotropy condition $p_r=p_t$, charged
fluid is a neutral fluid with a special kind of anisotropy, conformal
flatness means the constraint $\Psi _2=0$, uniform density constrains the
energy density to a function of time, anisotropy may be studied in more
detail by a linear equation among the pressures, which includes the cases
where one of them vanishes. In this way the relations among the particular
cases become much clearer due to the relations of all of them to the core.
Time and again the second order Emden-Fowler equations appear, which
sometimes may be integrated to first order and even to implicit integral
formulas for the solution, like Eqs (36, 50, 58). In many cases they are
expressed through elementary functions.

Imposing one constraint fixes the arbitrary potential of the general model
but up to functions of time or radius. Imposing a second constraint makes
the system overdetermined, yet special solutions still exist when the
arbitrary functions become constants. In total, we find many uncharged and
charged anisotropic solutions, all conformally flat solutions, a large class
of solutions with proportional or vanishing pressures and some uniform
density solutions. The general geodesic dust solutions are found explicitly
in all cases, building upon the LTB solution.

The present classification is based on the mass function formalism, which is
applicable only to metrics varying with time. Therefore the static case
should be studied on the base of the usual Einstein equations (2) which
simplify a lot in this case.

Much work remains to be done. One should classify as a next step the
expansion-free anisotropic spheres. Physically realistic solutions for
collapsing star models should be distinguished from the above classes, cases
and subcases with numerous arbitrary functions and constants of integration.
Junction conditions give additional constraints. It seems possible to
generalize the above constructions also to radiating anisotropic spheres,
which have attracted a lot of interest recently.

.

\end{document}